\documentclass[aps,pra,twocolumn,groupedaddress,amsmath,amssymb]{revtex4-1}

\usepackage{graphicx}
\usepackage{dcolumn}
\usepackage{bm}
\usepackage{amsmath}
\usepackage{color}
\usepackage{multirow}
\usepackage{threeparttable}
\usepackage{float}

\def\apj{ApJ}                 
\def\apjl{ApJ}                
\def\aap{Astron. \&  Astroph.}                
\def\araa{ARA\&A}             
\def\apjs{ApJ}                
\def\apss{ApSS}		
\def\mnras{MNRAS}             
\def\nat{Nature}              
\def\pra{Phys. Rev. A} 			
\def\pre{Phys. Rev. E} 			

\newcommand{\bra}[1]{\langle #1|}
\newcommand{\ket}[1]{|#1\rangle}

\begin{document}


\title{The carbon atom in intense magnetic fields}


\author{Anand Thirumalai}
\email[Corresponding author, SESE Exploration Postdoctoral Fellow, electronic address: ]{anand.thirumalai@asu.edu.}
\affiliation{School of Earth and Space Exploration, Arizona State University, Tempe, Arizona, USA, 85287}
\author{Steven J. Desch}
\email[]{steve.desch@asu.edu}
\affiliation{School of Earth and Space Exploration, Arizona State University, Tempe, Arizona, USA, 85287}
\author{Patrick Young}
\email[]{patrick.young.1@asu.edu}
\affiliation{School of Earth and Space Exploration, Arizona State University, Tempe, Arizona, USA, 85287}


\date{\today}

\begin{abstract}
The energy levels of the first few low-lying states of carbon in intense magnetic fields upwards of $\approx 10^7$~T are calculated in this study. We extend our previously employed pseudospectral approach for calculating the eigenstates of the carbon atom. We report data for the ground state and a low-lying state that are in good agreement with findings elsewhere, as well as new data for ten other states of the carbon atom that have not been investigated until now. It is seen that these hitherto uninvestigated states also become strongly bound with increasing magnetic field strengths. The data presented in this study are relevant for astrophysical applications, such as magnetized white dwarf and neutron star spectral analysis as well as opacity calculations and absorption features, including in the context of material accreting onto the surfaces of these compact objects. 
\end{abstract}

\maketitle

\section{\label{sec:intro}Introduction}

The study of atoms in magnetic fields of strength beyond the perturbative regime was largely motivated by the discovery of strong fields being present in white dwarf stars \cite{Kemp1970, Angel1978, Angel1981} and neutron stars \cite{Trumper1977,Trumper1978}. Pulsars, which are the most commonly observed neutron stars, harbor intense magnetic fields on the order of $10^{7}$ - $10^{9}$T \cite{Ruder94}. Magnetars \cite{DT1992}, which are strongly magnetized neutron stars, can have field strengths well in excess of $10^{9}$T. White dwarfs posess somewhat weaker but nevertheless still strong magnetic fields, with strengths $\sim10^{2}$ - $10^{5}$T \cite{Ruder94}. Even at these somewhat lower white dwarf field strengths, atomic structure is considerably altered from the low-field case, and a Zeeman-type perturbative treatment of the field \cite{Landau} is not possible. 

The need to calculate atomic structure in high magnetic fields has gained considerable impetus
in recent years. It is emerging from X-Ray observations that neutron stars atmospheres may 
contain mid-$Z$ neutral atoms, especially carbon \cite{Ho2009, Suleimanov2014}, cosmo-chemically one of the most important elements. Interpretation of the emergent spectra is hindered by the lack of atomic data pertinent to the extreme
environment of neutron star atmospheres. The presence of strong electric as well as magnetic fields have a profound influence on the emergent spectra altering the energy levels and ionization potentials and affecting ion population distributions at different energy levels. The ubiquitous effects of line broadening further complicate spectral analysis. It is remarkable that even for the simpler case of no electric field, and relatively weak (for a neutron 
star) magnetic field $B \sim 10^{6}$~T, photoionization edges and spectral lines differ significantly from the field-free case. A considerable amount of atomic data is therefore required for accurately interpreting the spectra of neutron stars. 

Similarly, observations of white dwarfs are also motivating study of atoms in intense magnetic fields. A sizable number of white dwarfs are highly magnetized, with magnetic fields around or in excess of $10^5$~T \cite[see][for a short review]{Tout2008}. It is also now emerging that about $25\%$ of white dwarfs are contaminated with mid-$Z$ atoms such as 
carbon, silicon, phosphorus and sulphur \cite{Barstow2014, Koester2014}. The existence of such contaminants in white dwarf atmospheres has been a surprise, because  stellar 
evolution models predict largely H or He atmospheres (DA or DB white dwarfs, respectively), with heavier species sinking on relatively short timescales $\sim10^2$ yr \cite[e.g.][]{Koester2014}. The heavier atoms (such as silicon, phosphorus and sulphur) are predominantly present in hotter white dwarfs where they are still radiatively levitated before submerging, although some observations reveal that even cooler white dwarfs show such contaminants \cite{Barstow2014}. Carbon meanwhile has been observed in a large variety of white dwarfs, both hot and cooler ones. To reconcile these observations, an exogenous source is therefore argued for, and it is becoming understood that white dwarfs often accrete the remnants of planetary systems. Such observations are being used to determine planetary compositions in these erstwhile systems \cite{Koester2014}. Recently, a DQ white dwarf (spectra distinguished by the presence of carbon lines), $\textrm{SDSS~J}142625.71+575218.3$ has been observed to harbor a magnetic field of strength $\sim 1.2 \times 10^{5}$~T \cite{Dufour2008}, further motivating the need for atomic data for carbon in intense magnetic fields, such as energy levels of different orbitals alongside electron densities, with data for oscillator strengths for bound-free and free-free transitions, to facilitate spectral analysis. 

An additional need for basic atomic data stems from the realization that in the atmospheres of magnetized white dwarfs and neutron stars, the atomic orbitals of adjacent atoms may bond via a new mechanism, the so-called perpendicular paramagnetic bonding, which can lead to strongly bound ${\rm H}_{2}$, ${\rm He}_{2}$ \cite[][]{Lange2012}, and possibly other species as well. In these highly magnetized astrophysical objects, even simple atoms behave completely differently from their terrestrial counterparts. It is such considerations that have motivated the current study. We present below a short review of the literature pertaining to this field of study. The reader is referred to a recent article \cite[][]{Thirumalai2014a} for a more detailed review of this area or research. 
 
Due of the impossibility of achieving such high magnetic field strengths in laboratory settings,
atomic data for high-$B$ atoms have traditionally been derived using modeling.
A variety of techniques have been used by various researchers since the 1970s, mostly applied to the hydrogen atom \cite{CK1972, Praddaude1972, SV1978, Friedrich1982, WR1980, RWRH1983, RWRH1984, Ivanov1988, VB2008, VB2002} and many recent studies of helium \cite{Proschel1982, Thurner1993, Ivanov1994, Jones1996, Jones1997, Jones1999, HH1998, MH2002, MH2007, Schmelcher2003, Schmelcher2003II, Schmelcher2002, Schmelcher2001, Schmelcher2000, Schmelcher1999, Schmelcher_helium2007, Ivanov1997, Ivanov2000, Wang2008} in strong magnetic fields. Studies have also been conducted for molecules and chains of both hydrogen and helium atoms relevant to neutron star magnetic fields \cite{Lai1992, Lai1993, Lai1996, Medin2006I, Medin2006II, Schmelcher97hyd, Schmelcher2000hyd, Schmelcher2001hyd}. Recent investigations by Thirumalai \& Heyl \cite{TH2009} using single-configuration Hartree-Fock (HF) theory \cite{Hartree} was seen to yield accurate upper bounds for the binding energies of hydrogen and helium in strong magnetic fields. A follow-up study \cite{HT2010}, obtained accurate binding energies for helium and lithium atoms in strong magnetic fields using a pseudospectral method. This approach was seen to be computationally far more economical than using the earlier finite-element based approach \cite{TH2009}. 

In contrast to the somewhat simpler two-electron systems, relatively little work exists in the literature for atoms with more than two electrons in strong magnetic fields. One of the first studies to investigate atoms in intense magnetic fields, in particular the iron atom, was by Flowers et al \cite{Flowers77} in 1977. This variational study extended the work due to the authors in Ref.~\cite{GK1975} and obtained binding energies of iron atoms and condensed matter in magnetic fields relevant to neutron stars. Errors in this study were later corrected by Muller \cite{Muller83}. Other methods such as density functional studies \cite{PBJones85_1, PBJones85_2} and the Thomas-Fermi-Dirac method \cite{Mueller71, Skjervold84} were also employed for estimating binding energies of atoms in intense magnetic fields. Recently, Medin \& Lai \cite{Medin2006I, Medin2006II} have also studied atoms and molecules and infinite chains of condensed matter in magnetic fields greater than $10^{8}$~T, using density-functional-theory. Mori et al \cite{MH2002, MH2007} have studied mid-Z atoms in strong to intense magnetic fields using perturbation theory as well, obtaining results consistent with previous findings.

The first comprehensive HF studies of atoms with more than two electrons were carried out by Neuhauser et al \cite{Neuhauser86, Neuhauser87} for magnetic fields greater than $10^{8}$~T, thus being directly relevant to neutron stars. Elsewhere, HF studies of atoms and molecules in intense magnetic fields were conducted by Demuer et al \cite{Demeur94}, with results consistent with previous findings. All of the above treatises, Refs.~\cite{Flowers77, GK1975, Muller83, PBJones85_1, PBJones85_2, Mueller71, Skjervold84, Neuhauser86, Neuhauser87, Demeur94}, concern themselves with magnetic fields in excess of $10^{8}$~T, well into the so-called intense magnetic field regime. At these field strengths, the interaction of the electron with the nucleus of the atom becomes progressively less dominant, in comparison to its interaction with the field itself. 

Various fully computational methods have been brought to bear on the case of atoms with more than two electrons in strong fields. One of the first studies to carry out a rigorous HF treatment of atoms with more than two electrons in strong or intermediate field strengths was Ref.~\cite{Jones1996}. Therein, they obtained estimates of the binding energies of a few low-lying states of lithium and carbon atoms, in low to strong magnetic fields. In recent years, Ivanov \cite{Ivanov1998} and Ivanov \& Schmelcher \cite{Ivanov1997, Schmelcher1999exch, Ivanov1999, Ivanov2000, Ivanov2001EPJD, Ivanov2001JPhB} have carried out detailed HF and post-HF studies of multi-electron atoms using a numerical mesh-method for solving the unrestricted HF equations \cite{Ivanov1997}. The special meshes were constructed so as to facilitate finite-difference calculations in a two-dimensional domain using carefully selected mesh node points \cite{Ivanov2001}. Using this method they were able to ascertain the binding energies of the first few low-lying states of low-Z atoms such as lithium, beryllium and mid-Z atoms such as boron and carbon etc. Al-Hujaj \& Schmelcher \cite{Schmelcher_lithium2004, Schmelcher_beryllium2004} adopting a full configuration-interaction method, using a gaussian basis for the electron wave functions \cite{Schmelcher2003, Schmelcher2003II, Schmelcher2002, Schmelcher2001, Schmelcher2000, Schmelcher1999, Schmelcher_helium2007}, obtained accurate estimates of the binding energies of lithium and beryllium atoms in strong magnetic fields. The sodium atom in a strong magnetic field has also been studied by Gonzalez-Ferez \& Schmelcher \cite{Schmelcher_sodium2003} obtaining estimates for the binding energies. Elsewhere, low lying states of the lithium atom have also been studied in strong magnetic fields using a configuration-interaction method, employing the so-called freezing full-core method both with \cite{Qiao2000} and without \cite{Guan2001} correlation between electrons. In recent years Engel and Wunner and co-workers  \cite{Engel2008, Engel2009, Schimeczek2013, Schimeczek2012, Klews2005} have computed accurate results for several atoms in magnetic fields relevant to neutron stars with a variety of techniques involving finite-element methods with B-splines both in the adiabatic approximation and beyond the adiabatic approximation with more than one Landau level. These highly accurate formulations employ a fast parallel Hartree-Fock-Roothan code, in which the electronic wave functions are solved for along the $z-$direction, with Landau orbitals (and combinations of more than one level in the latter studies) describing the remaining parts of the wave functions. Elsewhere, different \emph{ab initio} Quantum Monte-Carlo approaches \cite{Bucheler2007, Bucheler2008} have also been successfully employed to determine the ground states of atoms and ions in strong magnetic fields. Recently excited states of helium have also been computed quite accurately in intense magnetic fields using a fixed-phase Quantum Monte-Carlo approach \cite{Meyer2013}. 

Recently, in very comprehensive studies, \citet[][]{Schimeczek2012, Schimeczek2013} and \citet{Boblest2014} obtained accurate estimate of the ground state energies of atoms and ions up to $Z=26$, with only a few seconds worth of computing time for helium and heliumlike atoms. Such speeds are essential for coupling atomic structure codes with atmosphere models and spectral analysis codes for magnetized white dwarfs and neutron stars. While these investigations concerned themselves with the ground state, a recent study by Thirumalai \& Heyl \cite[][]{Thirumalai2014b} employed a fast pseudospectral approach for computing accurately the first-few low-lying states of helium and lithium in intense magnetic fields. They obtained data for two previously un-investigated states of lithium which were observed to become tightly bound with increasing magnetic field strength. By virtue of spectral convergence, the computational times of this approach were seen to be on the order of seconds for heliumlike systems, while maintaining accuracy. 

The current study extends the approach due to Thirumalai \& Heyl \cite[][]{Thirumalai2014b} to the carbon atom, investigating the first few low-lying states that become tightly bound in the limit of intense magnetic fields. The article is arranged as follows. In Sec.~\ref{sec:HF} we provide the basic governing equations, in Sec.~\ref{sec:numer} we briefly describe the numerical methodology adopted for solving the eigensystem. In Section~\ref{sec:results} we present and discuss the results. Finally in Sec.~\ref{sec:conclusion} we present conclusions and briefly describe avenues for further investigation. 

\section{\label{sec:HF}The HF Equations}

We begin with the generalized single-configuration HF equations for an atom with $N$ electrons and nuclear charge $Z$, in a magnetic field that is oriented along the $z$-direction \cite[][]{TH2009,Thirumalai2014b}. The single configuration HF equation can be written in cylindrical coordinates as, [where the length scale is in units of Bohr radii and the energy is scaled in units of Rydberg energy in the Coulomb potential of charge $Ze$; see below for definitions]
 \begin{widetext}
 \begin{eqnarray}
 \left[-\nabla^{2}_{i}(\rho_{i},z_{i})+\frac{m_{i}^{2}}{\rho_{i}^{2}}+2\beta_{Z}(m_{i}-1)
 +\beta_{Z}^{2}\rho_{i}^{2}
 -\frac{2}{r_i}\right]\psi_{i}\left(\rho_{i},z_{i}\right)
 +\frac{2}{Z}\sum_{j\neq
   i}\left[\Phi_{D}\psi_{i}(\rho_{i},z_{i})-\alpha_{E}\psi_{j}(\rho_{i},z_{i})\right]=\epsilon_{i}\psi_{i}\left(\rho_{i},z_{i}\right),\label{eq:1}
 \end{eqnarray}
 where \begin{math}i,j=1,2,3,...,N\end{math} and
 \begin{math}r_i=\sqrt{\rho_i^2+z_i^2}.\end{math} Please note that the three-dimensional momentum operator has been split into two parts: $\nabla_{i}^2(\rho_i,z_i)$ which are the $\rho-$ and $z-$ parts of the Laplacian; and $m_i^2/\rho_i^2$ which is the azimuthal part. The total Hartree-Fock energy of
 the state is given by
 \begin{equation}
 \varepsilon_{total}=\sum_{i}\epsilon_{i}-\frac{1}{2}\frac{2}{Z}\sum_{j\neq i}\left[\bra{\psi_{i}(\rho_{i},z_{i})}\Phi_{D}\ket{\psi_{i}(\rho_{i},z_{i})}-\bra{\psi_{i}(\rho_{i},z_{i})}\alpha_{E}\ket{\psi_{j}(\rho_{i},z_{i})}\right].\label{eq:2}
 \end{equation}
 The direct ($\Phi_D$) and exchange ($\alpha_E$) interactions are determined according to the method outlined in Ref.~\cite{TH2009}, as the solutions of the elliptic partial differential equations for the potentials given by
 \begin{equation}
 \nabla^{2}_{i}\Phi_{D}=-4\pi|\psi_{j}(\rho_{i},z_{i})|^{2}
 \label{eq:3}
 \end{equation}
 and
 \begin{equation}
 \left[
 \frac{1}{\rho_{i}}\frac{\partial}{\partial\rho_{i}}
 	\left(\rho_{i}\frac{\partial}{\partial\rho_{i}}\right)-\frac{(m_{i}-m_{j})^{2}}{\rho_{i}^{2}}+\frac{\partial^{2}}{\partial z_{i}^{2}}\right]\alpha_{E}(\rho_{i},z_{i})
 =-4\pi\psi_{j}^{*}(\rho_{i},z_{i})\psi_{i}(\rho_{i},z_{i}).
 \label{eq:4}
 \end{equation}
 \end{widetext}
Here $\psi_i$ and $\psi_j$ are the wave functions of the $i^\textrm{th}$ and $j^\textrm{th}$ electrons. The wave function of a given configuration of electronic orbitals is assumed to be given by a single Slater determinant as
 \begin{equation}
 \Phi=A_{N}\left(\tilde{\psi}_{1}, \tilde{\psi}_{2}, \tilde{\psi}_{3}, ..., \tilde{\psi}_{N-1},\tilde{\psi}_{N}\right),
 \label{eq:5}
 \end{equation}
 where \begin{math}A_{N}\end{math} is the anti-symmetrization operator. The individual electronic wave functions $\tilde{\psi}_{i}$ are given by
 \begin{equation}
 \tilde{\psi}_{i}= \psi_{i}(\rho_{i},z_{i})e^{im\phi_{i}}\chi_{i}(s_{i}),
 \label{eq:6}
 \end{equation}
 where \begin{math}i\end{math} labels each of the $N$ electrons. The two-dimensional single particle wave functions \begin{math}\psi_{i}(\rho_{i},z_{i})\end{math} are taken to be real functions. $\chi_{i}(s_{i})$ are the spin parts of the wave functions.

Integration with respect to the azimuthal coordinate, $\phi$, has been carried out, prior to writing the result in Eq.~(\ref{eq:1}) above. The contribution due to electron spin has also been averaged out \textit{a priori}. It is to be mentioned in this regard that in the current study we shall only be concerned with fully spin-polarised states (FSP); in other words all the electrons of the atom are assumed to be anti-aligned with the magnetic field. Such states have an exchange interaction between the electrons providing an extra coupling term in the HF equations, $\alpha_E$. Additionally, FSP states are seen to be the most tightly bound states in the intense field regime. The extension to partially spin-polarised configurations is easily achieved by eliminating the exchange term in the HF equations. In the current study, we have chosen to work in units of Bohr radii along with the definitions given below. 

The Bohr radius for an atom of nuclear charge $Z$ is given by $a_{B}/Z$, where $a_B=\hbar/\alpha m_e c$ is the Bohr radius of the hydrogen atom. The magnetic field strength parameter $\beta_Z$, is given by the expression $\beta_Z=B/(Z^2B_0)$, where $B_0$ is the critical field strength at which point the transition to the intense magnetic field regime occurs \cite{Ruder94}. This is defined as  $B_{0}=(2\alpha^{2}{m_{e}}^{2}c^{2})/(e\hbar) \approx 4.70108 \times 10^5$T. Thus, beyond a value of $\beta_Z\approx1$, the interaction of the electron with the nucleus becomes progressively less dominant as $\beta_Z$ increases. Based upon the above definition of $\beta_Z$, it is convenient to classify the field strength \cite{Jones1996} as low ($ \beta_Z \leq 10^{-3}$), intermediate, also called strong ($ 10^{-3} \leq \beta_Z \leq 1 $) and intense or high ($  1 \leq \beta_Z \leq \infty $). These definitions of the different magnetic field strength regimes are useful to remember when discussing the results in the latter part of this paper and for distinguishing between ``strong" and ``intense" magnetic field strengths. The current study concerns itself with the \emph{intense} magnetic field regime. 
 
The energy parameter of the $i^{th}$ electron is defined as $\epsilon_{i}=E_{i}/(Z^{2}E_{\infty})$, with $E_{\infty}=\frac{1}{2}\alpha^2m_e c^2$, the Rydberg energy of the hydrogen atom. For brevity we shall refer to the units of energy as $E_{Z,\infty}$, which should be remembered as the Rydberg energy in the Coulomb potential of charge $Ze$. The quantity $\alpha=e^2/(4\pi\epsilon_{0}\hbar c)\approx1/137$ is the fine structure constant. In the current study, all the physical constants were used in SI units, with the magnetic field $B$ in Tesla. Eq.~(\ref{eq:1}) represents the $N$-coupled Hartree-Fock equations in partial differential form for an $N$-electron system with nuclear charge $Z$. The equations are coupled through the exchange interaction term between the electrons and as such the system of equations is solved iteratively. In the following section we briefly describe the numerical methodology employed in the current study. For solving the system of partial differential equations we adopted a pseudospectral approach and utilized an atomic structure software package developed earlier \cite[][]{Thirumalai2014b}. 

\section{\label{sec:numer}Numerical Details}

The numerical solution of the coupled eigenvalue problem in Eq.~(\ref{eq:1}) proceeds via the so-called self consistent field (SCF) method due to Hartree \cite{Hartree}. First we find a solution to the hydrogenic problem, i.e., Eq.~(\ref{eq:1}) without the direct and exchange interactions. This yields ionic single electron hydrogenic wave functions in the Coulomb potential of charge $Ze$ forming the initial estimates for the HF iterations. Second, using these estimates, the elliptic partial differential equations for the direct and exchange interaction potentials in Eqs.~(\ref{eq:3}) and (\ref{eq:4}) are solved. With these potentials now obtained, the coupled HF problem including the direct and exchange interactions in Eq.~(\ref{eq:1}) is solved as an eigensystem. The exchange interactions that couple the equations are expressed using wave-functions from the previous iteration to solve the eigenvalue problem for each electron \cite{Slater1951}. The eigenvalues obtained are the individual particle energies $\epsilon_i$ and the normalized eigenvectors are the wave functions, $\psi_i(\rho_i,z_i)$. The SCF iterations then proceed with the updated electron wave functions and the steps from the second step described above, are repeated until convergence.


For transforming the partial differential equations into algebraic ones, we follow the domain discretization procedure described in detail in Ref.~\cite[][]{Thirumalai2014b}. The salient points are given below in brief. As a result of azimuthal symmetry of the problem, and parity with respect to the $z=0$ plane, it is sufficient to restrict the physical domain of the problem  \cite{Ruder94, TH2009, Thirumalai2014b} to $ 0  \leqslant \rho, z \leqslant \infty$. However, for making the problem numerically tractable, instead of using the above semi-infinite domain, we instead solve the problem in a finite albeit sufficiently large domain of size $\rho_{\textrm{max}} \times z_{\textrm{max}}$. This finite domain is then mapped using a suitable transformation (see below) to the domain $[-1,1]$, and Chebyshev-Lobatto spectral collocation points are then located on this latter compactified domain \cite{Trefethen}. Thereafter, a Chebyshev pseudospectral method can be employed for representing the differential operators and functions in this transformed domain. However, domain truncation can introduce a confinement energy as an artifact of the numerical procedure, artificially increasing the binding energy of the electron \cite{Thirumalai2014b}. This is mitigated by using a sequence of domains of increasing sizes, obtaining a converged result in the limit of the computational domain approaching the size of the physical domain of the problem \cite{Thirumalai2014b}. 

In our computations, the size of the computational domain $\rho_{\textrm{max}}$ and $z_{\textrm{max}}$ (in units of Bohr radii) are given by
 \begin{equation}
 \rho_{\textrm{max}} ~,~ z_{\textrm{max}} = \frac{100 \eta}{1+\textrm{log}_{10}(\beta_Z)},
 \label{eq:compactification}
 \end{equation} 
where $\eta=\frac{1}{4}, \frac{1}{2}, 1, 2$ is a scaling factor used for setting up computations in a sequence of increasing domain sizes. The effect of the logarithmic term $\textrm{log}_{10}(\beta_Z)$ in the denominator is that it naturally makes the domain larger or smaller, depending on whether $\beta_Z < 1$ or $\beta_Z > 1$, respectively. With the maximum domain size thus defined, we can then compactify the finite domain $[0,\rho_{\textrm{max}}] \otimes [0,z_{\textrm{max}}]$ to $[-1,1] \otimes [-1,1]$ with the transformation,
 \begin{equation}
 x = \textrm{log}_{10} (1 + \rho \alpha_{\rho} ) - 1
 \label{eq:transform_rho}
 \end{equation}
 and
 \begin{equation}
 y = \textrm{log}_{10} (1 + z \alpha_{z} ) - 1,
 \label{eq:transform_z}
 \end{equation}
where $\alpha_{\rho}=99/\rho_{\textrm{max}}$ and $\alpha_{z}=99/z_{\textrm{max}}$. Note that in our calculations we employed a square domain for achieving the best possible internally consistent convergence. Therefore in our work $\rho_{\textrm{max}} = z_{\textrm{max}}$ and therefore $\alpha_\rho = \alpha_z \equiv \alpha$, but the possibility remains for using different sizes and scalings in the two orthogonal directions for optimizing computational effort, particularly in the intense field regime. 

In order to obtain a covered solution within any given domain size, we employed six different levels of mesh refinement using $N=21,31,41,51,61~\textrm{and}~71$ Chebyshev collocation points in each of the two orthogonal directions.

Utilizing a pseudospectral approach for discretization results in a sparse matrix for the coupled eigenvalue problem  \cite{Thirumalai2014b}. Therefore we employ the widely used sparse matrix generalized eigensystem solver ARPACK, which utilizes the implicitly restarted Arnoldi method (IRAM) \cite{Arnoldi1951,Sorensen1992,ARPACK,Saad1984} for solution. The key advantage is that since the Hamiltonian matrix that we are solving only has a few bound state solutions, employing IRAM with the shift-invert algorithm \cite{ARPACK} for computing only a portion of the spectrum saves considerable computational effort. 

It was found that generating a Krylov subspace with about $50$ to $250$ basis vectors was sufficient for determining around $15$ to $100$ eigenvalues in the vicinity of a given shift ($\sigma$). Runs were carried out for different values of the magnetic field strength parameter $ \beta_Z $, in the range $ 0.7 \stackrel{_<}{_\sim} \beta_Z \leq
   250 $, for the cylindrical pseudospectral code. A typical
 tolerance of around $10^{-10}$ was employed for the internal errors of
 ARPACK. It was observed during our runs that fast convergence was achieved; within about $3-6$ HF iterations. A convergence criterion for the HF iterations was employed wherein the difference between the HF energies for two consecutive iterations was tested. Typically, a tolerance on the order of $10^{-6} E_{Z,\infty}$ was employed. Once the HF iterations attained convergence for a given level of mesh refinement, the total energy of the Hartree-Fock state under consideration is reported according to Eq.~\ref{eq:2}. 
 
\section{\label{sec:results} Results and Discussion}

Using the atomic structure software package developed for an earlier study \cite{Thirumalai2014b}, we carried out computations for several FSP states of the neutral carbon atom in intense magnetic fields. After applying the extensive convergence conditions to the computations as described in Ref.~\cite{Thirumalai2014b}, we arrived at estimates of the binding energies for the $12$ tightly bound states in the intense field regime. Among these, only two states have been investigated earlier; therefore the majority of the data presented herein aims to complement the already available data.

The states that were considered in this study are labelled using both the field-free and strong-field notations for the sake of clarity; these can be found in Table~\ref{tab:table1}, which lists the different states of carbon. In the presence of a magnetic field states can be characterized using the notation $^{2S+1}\textrm{M}^{\pi_z}$, where $\textrm{M}=\Sigma_i m_i$ is the total $z-$ component of angular momentum. The summation is over all the electrons in the atom. This then forms a manifold within which different sub-spaces exist. The spin multiplicity is given in the usual way as $2S+1$. Finally, the $z-$parity of the state is indicated using $\pi_z= \pm 1$, indicating positive or negative parity. We studied $12$ tightly bound states of carbon, $6$ in each $z-$parity subspace, in the intense magnetic field regime ($\beta_Z \stackrel{_>}{_\sim} 1$). Within a given parity sub-space, typically there are crossovers that occur as the magnetic field is reduced; the reader is referred to \citet{Ivanov1999,Ivanov2000} for excellent data and discussions regarding ground state crossovers. A recent study by \citet{Boblest2014} also represents one of the most comprehensive discussions with regard to transitions concerning the ground states of atoms up to $Z=26$. The current work adds to the available atomic data by investigating sates of the carbon atom not considered in these studies and reports on the binding energies of several low-lying states in the intense field regime.
 \begin{table}[h]
 \centering
 \caption{The different states of carbon considered in this study, listed using both intense-field and field-free notation.}
 \begin{tabular}{c@{\hspace{3mm}}c@{\hspace{5mm}}}
 \hline
 \hline
Intense-field ~ & ~ Field-free \\
 \hline
$^7(-15)^+$ & $1s_02p_{-1}3d_{-2}4f_{-3}5g_{-4}6h_{-5}$ \\
$^7(-15)^-$ & $1s_02p_{-1}3d_{-2}4f_{-3}5g_{-4}7i_{-5}$ \\
& \\
$^7(-14)^+$ & $1s_02s_{0}3d_{-2}4f_{-3}5g_{-4}6h_{-5}$ \\
$^7(-14)^-$ & $1s_02p_{-1}3d_{-1}4f_{-3}5g_{-4}6h_{-5}$ \\
& \\
$^7(-13)^+$ & $1s_02s_{0}2p_{-1}4f_{-3}5g_{-4}6h_{-5}$ \\
$^7(-13)^-$ & $1s_02p_{0}2p_{-1}4f_{-3}5g_{-4}6h_{-5}$ \\
& \\
$^7(-12)^+$ &  $1s_02s_{0}2p_{-1}3d_{-2}5g_{-4}6h_{-5}$ \\
$^7(-12)^-$ & $1s_02p_{0}2p_{-1}3d_{-2}5g_{-4}6h_{-5}$ \\
& \\
$^7(-11)^+$ &  $1s_02s_{0}2p_{-1}3d_{-2}4f_{-3}6h_{-5}$ \\
$^7(-11)^-$ &  $1s_02p_{0}2p_{-1}3d_{-2}4f_{-3}6h_{-5}$ \\
& \\
$^7(-10)^+$ & $1s_02s_{0}2p_{-1}3d_{-2}4f_{-3}5g_{-4}$ \\
$^7(-10)^-$ & $1s_02p_{0}2p_{-1}3d_{-2}4f_{-3}5g_{-4}$ \\
 \hline
 \hline
 \end{tabular}
 \label{tab:table1}
 \end{table}

\subsection{\label{sec:positive_parity}The positive parity ($\pi_z=+1$) subspace}

For the states of carbon listed in Table~\ref{tab:table1}, eigenvalues were determined using the numerical method described in Section~\ref{sec:numer} (see Ref.~\cite{Thirumalai2014b} for more details). We began with the lowest value of the domain scaling parameter $\eta=1/4$. This yielded a domain with dimensions given according to Eq.~(\ref{eq:compactification}), and this domain size depends on $\beta_Z$. HF energies were then calculated using up to six different levels of mesh refinement in the domain. This enabled us to extrapolate the results to the limit of infinitely fine mesh, for a given domain size. We observed exponential convergence, characteristic of spectral methods, wherein the errors diminish exponentially with mesh refinement. We employed an exponential function of the form $ae^{bx}+ce^{dx}$ for extrapolating the binding energies to the limit of infinitely fine mesh. A Levenberg-Marquardt optimization algorithm \cite{NR1992} was employed for this purpose. The errors associated with the extrapolation procedure were typically on the order of $10^{-4}$ to $10^{-6} E_{Z,\infty}$ with a normalized $R-$squared value typically $>0.999$ for the interpolating function employed. However, at the upper end of the intense magnetic field regime, we noticed slight loss of accuracy as the states become tightly bound, and for $\beta_Z \stackrel{_>}{_\sim} 200$ the extrapolation procedure had an error on the order of few times $10^{-4} E_{Z,\infty}$ with a normalized $R-$squared of $\approx 0.98$ on average. For the extrapolation to infinitely fine mesh, the average area per unit grid size in the domain ($A_E \approx \rho_{\textrm{max}} z_{\textrm{max}}/N^2$), was taken as the independent variable and the energies extrapolated to the limit of $A_E \rightarrow 0$, corresponding to infinitely fine mesh.

This procedure was repeated as the domain was rescaled to larger and larger values, corresponding to $\eta=1/2, 1, 2$. Then, using the extrapolated values of the HF energy corresponding to infinitely fine mesh for each of the four domain sizes, a subsequent extrapolated value of the the HF energy ($E_{HF}$) was obtained, in the limit of the domain size approaching infinity. These are then the converged $E_{HF}$ values reported in Tables~\ref{tab:carbon_positive} and \ref{tab:carbon_negative}. We employed an extrapolating function of the form $ax^{1/2}+b$, with a Levenberg-Marquardt optimization method \cite{NR1992}. The ordinates in this case were the four different converged HF energies in the limit of infinitely fine mesh in each of the four different domains, and the abscissae were the inverse domain areas, i.e. $(\rho_{\textrm{max}} z_{\textrm{max}})^{-1}$. Thus, extrapolating to zero inverse area corresponding to an infinite domain size yields the final converged HF energy, and mitigates errors arising due to domain truncation \cite{Thirumalai2014b}. The error in the extrapolation to the limit of an infinite domain size was on the order of $10^{-5}$ to $10^{-6} E_{Z,\infty}$ with a normalized $R-$squared value of $>0.999$ for the interpolating function employed. Again at the upper end of magnetic field strengths ($\beta_Z \stackrel{_>}{_\sim} 200$) we noticed a slight loss of accuracy, with the extrapolation errors increasing to the level of a few times $10^{-4} E_{Z,\infty}$.
\begin{widetext}
 \begin{table}[H]
 \centering
 \caption{Absolute value of the binding energies of the positive parity states of carbon. Energies are in units of Rydberg energies in the Coulomb potential of nuclear charge $Z=6$ for carbon. Accurate data from other work is also provided for comparison. $\beta_Z=\gamma/2Z^2$. The values given in parentheses are the maximal fitting errors at the fourth decimal place.}
 \begin{threeparttable}
 \begin{tabular}{c@{\hspace{3mm}}c@{\hspace{3mm}}c@{\hspace{3mm}}c@{\hspace{3mm}}c@{\hspace{3mm}}c@{\hspace{3mm}}c@{\hspace{3mm}}c@{\hspace{3mm}}}
 \hline
 \hline
 & \multicolumn{2}{c}{$^7(-15)^+$} & \multicolumn{1}{c}{$^7(-14)^+$} & \multicolumn{1}{c}{$^7(-13)^+$} & \multicolumn{1}{c}{$^7(-12)^+$} & \multicolumn{1}{c}{$^7(-11)^+$} & \multicolumn{1}{c}{$^7(-10)^+$}\\ 
 \cline{2-2} \cline{3-4} \cline {5-8}\\
  $\beta_Z$ & Present work & Other work & Present work & Present work  & Present work & Present work & Present work\\
 \hline
 0.5909   &   3.7898(0)        &    3.7586\tnote{b}    &      &    &    &    &  \\
  0.6944  &   4.0165(1)        &    3.9794\tnote{a}    & 3.5545(1)     & 3.7294(1)   &  3.8084(0)  &  3.8592(2)  &3.9078(1)  \\
 1.0000   &   4.5889(1)        &                       & 4.0546(1)     & 4.2520(0)   &  4.3432(1)  &  4.4020(2)  &4.4583(1)  \\
 1.3889   &   5.1824(3)        &    5.1364\tnote{a}    & 4.5754(2)     & 4.7957(1)   &  4.8991(1)  &  4.9659(1)  &5.0301(1)  \\
 2.0000   &   5.9372(1)        &                       & 5.2398(4)     & 5.4887(2)   &  5.6071(1)  &  5.6840(1)  &5.7576(1)  \\
 2.7778   &   6.7127(1)        &    6.6563\tnote{a}    & 5.9245(2)     & 6.2018(2)   &  6.3355(3)  &  6.4223(2)  &6.5057(1)  \\
 2.9544   &   6.8692(1)        &    6.8213\tnote{b}    &      &    &    &    &  \\
 5.0000   &   8.3560(2)        &                       & 7.3771(2)     & 7.7130(2)   &  7.8780(1)  &  7.9859(1)  &8.0890(0)  \\
 5.9088   &   8.8894(1)        &    8.8339\tnote{b}    &      &    &    &    &  \\
 6.9444   &   9.4354(1)        &    9.3625\tnote{a}    &               &             &             &             &           \\
 7.0000   &   9.4632(1)        &                       & 8.3590(7)     & 8.7331(1)   &  8.9184(1)  &  9.0399(2)  &9.1561(0)  \\
10.0000   &   10.7851(2)       &                       & 9.5322(1)     & 9.9517(0)   & 10.1609(0)  & 10.2987(0)  &10.4307(4) \\
13.8889   &   12.1511(1)       &    12.0634\tnote{a}   & 10.7462(4)    & 11.2112(0)  & 11.4450(1)  & 11.5989(0)  &11.7461(0) \\
20.0000   &   13.8493(1)       &                       & 12.2583(1)    & 12.7780(2)  & 13.0411(2)  & 13.2152(0)  &13.3813(1) \\
25.0000   &   14.9905(1)       &                       & 13.2754(1)    & 13.8306(8)  & 14.1140(1)  & 14.3011(0)  &14.4798(1) \\
27.7778   &   15.5577(1)       &    15.4534\tnote{a}   & 13.7813(3)    & 14.3551(4)  & 14.6471(2)  & 14.8409(0)  &15.0259(0) \\
29.5440   &   15.8984(0)       &    15.8263\tnote{b}    &      &    &    &    &  \\
50.0000   &   19.0832(1)       &                       & 16.9279(1)    & 17.6089(9)  & 17.9602(0)  & 18.1937(2)  &18.4170(4) \\
69.4444   &   21.3415(2)       &    21.2117\tnote{a}   & 18.9470(2)    & 19.6950(3)  & 20.0822(1)  & 20.3405(2)  &20.5867(2) \\
100.0000  &   24.1131(4)       &                       & 21.4272(4)    & 22.2547(4)  & 22.6856(2)  & 22.9737(4)  &23.2492(1) \\
138.8889  &   26.8673(3)       &    26.7153\tnote{a}   &               &             &             &             &           \\
200.0000  &   30.2271(6)       &                       & 26.9070(22)   & 27.9012(33) & 28.4255(25) & 28.7783(20) &29.1161(1) \\
250.0000  &   32.4533(16)      &                       & 28.9040(15)   & 29.9575(16) & 30.5141(15) & 30.8902(6)  &31.2499(20)\\
   \hline 
 \hline
 \end{tabular}
 \begin{tablenotes}
        \item[a] Ref.~\cite{Ivanov2000}
        \item[b] Ref.~\cite{Schimeczek2013}
 \end{tablenotes}
 \end{threeparttable}
 \label{tab:carbon_positive}
 \end{table}
 \end{widetext}
It can be seen upon examining the data in Table~\ref{tab:carbon_positive} that only one FSP positive parity state of had been investigated in the intense field regime. This is the state $^7(-15)^+$ that becomes tightly bound, and is the ground state of the carbon atom, in the range of magnetic field strengths investigated in this study. It can be seen that the fully converged results obtained in the current study for this state are in good agreement with values obtained elsewhere, given that the current study is a single configuration calculation. Over the entire range of magnetic field strengths investigated, our estimates of the binding energies agree with estimates elsewhere \cite{Ivanov2000, Schimeczek2013} to on average $\Delta \approx 0.75\%$, for the states $^7(-15)^+$. We noticed a slight loss of accuracy of the cylindrical pseudospectral method in the lower magnetic field regime ($\beta_Z \stackrel{_<}{_\sim}$ 1) wherein the cylindrical code (and the extrapolation method described above) maintained accuracy to within $10^{-5}$ to $10^{-4}E_{Z,\infty}$. There was also a similar loss of accuracy at the upper end of intense field regime as well, where the electron orbital geometries become severely anisotropic ($\beta_Z \stackrel{_>}{_\sim} 200$). Binding energy data for five hitherto un-investigated FSP positive parity states is also provided in Table~\ref{tab:carbon_positive}. Within a given $M-\pi$ sub-space, we only considered a single state. It is therefore entirely possible that other states within this subspace have crossovers in the intense field and become tightly bound as well. This would require a detailed investigation of all the different states that can comprise a given $M-\pi$ sub-space. Such an investigation is left for a future undertaking, with a cautionary reminder to the reader that other states within a given sub-space apart from the ones listed here, may be important as well from a spectroscopic viewpoint. 
 
 \subsection{\label{sec:positive_parity}The negative parity ($\pi_z=-1$) subspace}
 
 \begin{widetext}
  \begin{table}[H]
 \centering
 \caption{Absolute value of the binding energies of the negative parity states of carbon. Energies are in units of Rydberg energies in the Coulomb potential of nuclear charge $Z=6$ for carbon. Accurate data from other work is also provided for comparison. $\beta_Z=\gamma/2Z^2$. The values given in parentheses are the maximal fitting errors at the fourth decimal place.}
 \begin{threeparttable}
\begin{tabular}{c@{\hspace{3mm}}c@{\hspace{3mm}}c@{\hspace{3mm}}c@{\hspace{3mm}}c@{\hspace{3mm}}c@{\hspace{3mm}}c@{\hspace{3mm}}c@{\hspace{3mm}}}
 \hline
 \hline
 & \multicolumn{1}{c}{$^7(-15)^-$}  & \multicolumn{1}{c}{$^7(-14)^-$} & \multicolumn{1}{c}{$^7(-13)^-$} & \multicolumn{1}{c}{$^7(-12)^-$} & \multicolumn{1}{c}{$^7(-11)^-$} &  \multicolumn{2}{c}{$^7(-10)^-$} \\ 
 \cline{2-2} \cline{3-4} \cline {5-8}\\
  $\beta_Z$ & Present work       & Present work & Present work & Present work & Present work & Present work         & Other work              \\
 \hline
  0.6944    &   3.9117(1)        &  3.7567(2)   &  3.7907(2)          &  3.8658(3)   & 3.9123(2)   & 3.9568(4)            & 3.9177\tnote{a}         \\
 1.0000     &   4.4630(0)        &  4.2766(6)   &  4.3028(2)          &  4.3903(3)   & 4.4451(2)   & 4.4979(4)            &                         \\
 1.3889     &   5.0346(1)        &  4.8156(2)   &  4.8354(3)          &  4.9355(3)   & 4.9991(2)   & 5.0601(3)            & 5.0153\tnote{a}         \\
 2.0000     &   5.7617(1)        &  5.5036(2)   &  5.5161(3)          &  5.6320(2)   & 5.7063(2)   & 5.7778(4)            &                         \\
 2.7778     &   6.5090(0)        &  6.2122(1)   &  6.2194(4)          &  6.3511(1)   & 6.4366(4)   & 6.5184(2)            & 6.4671\tnote{a}         \\
 5.0000     &   8.0927(0)        &  7.7195(1)   &  7.7209(1)          &  7.8854(3)   & 7.9926(2)   & 8.0954(2)            &                         \\
 6.9444     &                    &              &                     &              &             & 9.1338(2)            & 9.0672\tnote{a}         \\
 7.0000     &   9.1593(2)        &  8.7376(1)   &  8.7379(1)          &  8.9229(2)   & 9.0444(1)   & 9.1605(2)            &                         \\
10.0000     &   10.4312(1)       &  9.9535(2)   &  9.9534(1)          &  10.1625(1)  & 10.3001(2)  & 10.4316(2)           &                         \\
13.8889     &   11.7465(2)       &  11.2121(1)  &  11.2117(1)         &  11.4452(1)  & 11.5994(1)  & 11.7465(3)           & 11.6656\tnote{a}        \\
20.0000     &   13.3811(1)       &  12.7781(2)  &  12.7776(2)         &  13.0406(1)  & 13.2148(1)  & 13.3810(2)           &                         \\
25.0000     &   14.4793(2)       &  13.8314(7)  &  13.8305(1)         &  14.1132(2)  & 14.3004(2)  & 14.4793(1)           &                         \\
27.7778     &   15.0252(1)       &  14.3548(6)  &  14.3539(0)         &  14.6461(1)  & 14.8400(1)  & 15.0250(1)           & 14.9284\tnote{a}        \\
50.0000     &   18.4151(1)       &  17.6087(5)  &  17.6026(1)         &  17.9526(7)  & 18.1867(4)  & 18.4154(1)           &                         \\
69.4444     &   20.5850(0)       &  19.6006(16) &  19.6872(2)         &  20.0743(2)  & 20.3328(3)  & 20.5790(4)           & 20.4500\tnote{a}        \\
100.0000    &   23.2470(5)       &  22.2553(11) &  22.2523(3)         &  22.6831(3)  & 22.9713(4)  & 23.2522(5)           &                         \\
138.8889    &                    &              &                     &              &  	    & 25.8961(8)           & 25.7611\tnote{a}        \\
200.0000    &   29.1191(8)       &  27.9131(15) &  27.8991(22)        &  28.4235(25) & 28.8014(13) & 29.1198(8)           &                         \\
250.0000    &   31.2522(9)       &  29.9686(54) &  29.9602(68)        &  30.5110(15) & 30.8871(6)  & 31.2537(9)           &                         \\
   \hline                                                     
 \hline                              
 \end{tabular}
 \begin{tablenotes}
        \item[a] Ref.~\cite{Ivanov2000}
 \end{tablenotes}
 \end{threeparttable}
 \label{tab:carbon_negative}
 \end{table}

\end{widetext}

We investigated $6$ FSP negative parity states of the carbon atom in intense magnetic fields. We have provided data for the binding energies of these states in Table~\ref{tab:carbon_negative}. Of these, only the state $^7(-10)^-$ has been investigated earlier in the literature \cite{Ivanov2000}. The data of the current computation are seen to be in agreement with their results to on average $\approx 0.74\%$. It can also be seen that the state $^7(-15)^-$ is also a tightly bound state with the binding energies of this state being nearly equal to those of $^7(-10)^-$, making them the two most tightly bound states of the negative parity sub-space. We noticed that at both the low- and the high-end of the intense field regime considered here, there was a slight loss of accuracy, this can be seen in the slightly larger errors reported in the parentheses. Once more however, we would like to remind the reader that in any give $M-\pi$ sub-space, we have only investigated a single state; other configurations in the sub-space would need to be investigated before determining the ordering of states according to binding energy in a given $M-\pi$ sub-space, as well as answering the important question regarding crossovers. We again note that such an undertaking is beyond the scope of the current investigation whose aim is to merely complement the data in the literature for the carbon atom in intense magnetic fields, by providing data for hitherto un-investigated states that also become tightly bound with increasing magnetic fields. 

It can also be seen that binding energies of the different states shown in Tables~\ref{tab:carbon_positive} and \ref{tab:carbon_negative} are fairly close together, particularly at the higher end of the intense field regime, even with the handful of states considered here. This would have an impact on transitions probabilities wherein many transition probabilities between states may be nearly equally likely. This would also affect the emergent spectra wherein several lines may be rather close together. This effect may become more pronounced should other states in the different $M-\pi$ sub-spaces be investigated as well. 

\section{\label{sec:conclusion} Conclusion}

In the current study we have investigated the carbon atom in intense magnetic fields employing a two-dimensional single-configuration Hartree-Fock approach with a pseudospectral method of solution. We employed an atomic structure software package that was developed earlier \cite{Thirumalai2014b} for this purpose. 

We presented data for twelve tightly bound FSP states of carbon, six in each parity sub-space. Of these, ten of the states have not been investigated before. Where available, the data of the current computation for certain states were seen to be in good agreement with findings elsewhere.  

The pseudospectral atomic structure software employed in this investigation also has certain limitations. First, computations are currently required to be carried out in a sequence of increasing finite domain sizes, so that a converged result for the binding energy may be obtained in the limit of the domain size becoming infinite. This adds a layer of computational complexity. We have discussed in Ref.~\cite{Thirumalai2014b} a possible way to circumvent this, in essence by monitoring the wave functions at the outer edges of the domain and requiring their values to fall below a certain threshold, while varying the domain size. While this may not be straightforward to implement within the framework of a pseudospectral approach, it would nevertheless make the computation more streamlined if implemented. Second,  the current work does not include relativistic corrections to the energies. For the magnetic-field strengths considered herein, the relativistic corrections to the energies were estimated using the scaling formula in Ref~\cite{Poszwa2004}. Their results for the hydrogen atom were used for this purpose and the corrections were estimated to be on the order of $10^{-6}E_{Z,\infty}$. This was seen to be smaller than the numerical errors arising from convergence of the entire numerical method including
the extrapolation to the limit of a semi-infinite domain. Thus, relativistic corrections are important however it was not possible to account for them accurately in the current study. Moreover, as the magnetic field strength increases in the intense magnetic field regime, effects due to finite nuclear mass become relevant. In the current study, the mass of the nucleus is assumed to be infinite, and as such we have not carried out a suitable correction. One way to account for the finite nuclear mass is to employ a scaling relationship wherein the energies determined at a certain magnetic field strength $\beta_Z$ for an infinite nuclear mass, would be related to the corresponding binding energies for a finite nuclear mass at a different value of the magnetic field strength $\tilde{\beta}_Z$ \cite{Schmelcher1999}. 

Finally, and perhaps the most important, is the fact that the current study is only a single-configuration calculation for a system that has six electrons. Therefore, the effects of electron correlation are of great importance and, if included, would yield much more accurate results than those given here. The current $2-$D wave functions computed in this study could form the initial estimates for $2-$D configuration-interaction or multi-configuration calculations. We leave this much larger undertaking for a future endeavor. 

In summary, the current investigation considerably extends the currently available data in the literature for the carbon atom in intense magnetic fields. We would however like to remind the reader that several more states would need to be computed within the different $M-\pi$ sub-spaces for delineating the full energy landscape of the carbon atom. 

\begin{acknowledgments}
Calculations and the majority of the code development were performed on computing equipment purchased with funds from NSF grant
1148502 (PI John Shumway). A part of the code development was also carried out on computing infrastructure purchase with funds from the Canadian Foundation for Innovation and the British Columbia Knowledge Development Fund. AT gratefully acknowledges support from Arizona State University's School of Earth and Space Exploration Postdoctoral Fellowship.
\end{acknowledgments}


\bibliography{PaperIV}


\end{document}